\documentclass[a4paper,conference]{IEEEtran}
\IEEEoverridecommandlockouts

\newcommand\hmmax{0}
\newcommand\bmmax{0}

\usepackage{graphicx}
\usepackage{tikz}
\usepackage{pgfplots}
\usepackage{xcolor}
\usepackage{color, colortbl}
\usepackage{amsmath}
\usepackage{bbm}
\usepackage{amsfonts, amssymb}
\usepackage{mathtools}
\usepackage{cite}
\usepackage[nolist]{acronym}

\usepackage{booktabs} 		
\usepackage{diagbox}		
\usepackage{upgreek}
\usepackage{bbm}
\usepackage{Files/bm}
\usepackage{Files/winsnotation}
\usepackage{Files/Symbols_OPL}
\usepackage{Files/LVcolours} 

\usepackage[ruled,vlined,noresetcount]{algorithm2e}
\usepackage{setspace}	 	
\usepackage{comment}

\usepackage{array}
\newcolumntype{C}[1]{>{\centering\arraybackslash}p{#1}}

\makeatletter
\def\endthebibliography{%
  \def\@noitemerr{\@latex@warning{Empty `thebibliography' environment}}%
  \endlist
}
\makeatother
\IEEEoverridecommandlockouts
\usepackage{amsthm}
\theoremstyle{definition}

\newcommand{\rmt}t
\newcommand{\rmr}r

\pgfplotsset{compat=1.17}

\begin{document}

\title{ Impact of Interference Subtraction on Grant-Free Multiple Access with Massive MIMO}

\author{%
  \IEEEauthorblockN{Lorenzo Valentini, Alberto Faedi, Marco Chiani, Enrico Paolini}
  \IEEEauthorblockA{CNIT/WiLab, DEI, University of Bologna, Italy\\
  	Email: \{lorenzo.valentini13, alberto.faedi6, marco.chiani, e.paolini\}@unibo.it }
}

\maketitle 

\begin{acronym}
\small
\acro{ACK}{acknowledgement}
\acro{AWGN}{additive white Gaussian noise}
\acro{BCH}{Bose–Chaudhuri–Hocquenghem}
\acro{BS}{base station}
\acro{CDF}{cumulative distribution function}
\acro{CRA}{coded random access}
\acro{CRC}{cyclic redundancy check}
\acro{CRDSA}{contention resolution diversity slotted ALOHA}
\acro{CSA}{coded slotted ALOHA}
\acro{eMBB}{enhanced mobile broad-band}
\acro{FER}{frame error rate}
\acro{IFSC}{intra-frame spatial coupling}
\acro{i.i.d.}{independent and identically distributed}
\acro{IRSA}{irregular repetition slotted ALOHA}
\acro{LDPC}{low-density parity-check}
\acro{LOS}{line of sight}
\acro{MAC}{medium access control}
\acro{MIMO}{multiple input multiple output}
\acro{ML}{maximum likelihood}
\acro{MMA}{massive multiple access}
\acro{mMTC}{massive machine-type communication}
\acro{MPR}{multi-packet reception}
\acro{MRC}{maximal ratio combining}
\acro{PAB}{payload aided based}
\acro{PDF}{probability density function}
\acro{PHY}{physical}
\acro{PLR}{packet loss rate}
\acro{PMF}{probability mass function}
\acro{PRCE}{perfect replica channel estimation}
\acro{QPSK}{quadrature phase-shift keying}
\acro{RF}{radio-frequency}
\acro{SC}{spatial coupling}
\acro{SIC}{successive interference cancellation}
\acro{SIS}{successive interference subtraction}
\acro{SNB}{squared norm based}
\acro{SNR}{signal-to-noise ratio}
\acro{URLLC}{ultra-reliable and low-latency communication}
\end{acronym}
\setcounter{page}{1}

\begin{abstract}
The design of highly scalable multiple access schemes is a main challenge in the evolution towards future massive machine-type communications, where reliability and latency constraints must be ensured to a large number of uncoordinated devices. In this scenario, \ac{CRA} schemes, where \acl{SIC} algorithms allow large improvements with respect to classical random access protocols, have recently attracted an increasing interest. Impressive performance can be potentially obtained by combining \ac{CRA} with massive \ac{MIMO}. In this paper we provide an analysis of such schemes focusing on the effects of imperfect channel estimation on \acl{SIC}. Based on the analysis we then propose an innovative signal processing algorithm for \ac{CRA} in massive \ac{MIMO} systems.
\end{abstract}


\section{Introduction}


Next generation \ac{MMA} protocols should be designed to achieve very high scalability (number of simultaneously active users the system can support) in presence of reliability and latency constraints \cite{Hasan2013:random,Liu2018:sparse,Chen2020:massive,Gui2020:6G,Kalalas2020:massive,Pokhrel2020:Towards}. In this respect, grant-free multiple access schemes have gained an increasing interest, owing to their capability to substantially reduce control signalling for connection establishment, which is beneficial in terms of both scalability and latency.
Examples of grant-free schemes are the ones recently proposed in \cite{Liu2018:massivePt1,Sor2018:coded,Fengler2019:grant-free,Han2020:grant-free,choi2020:grantfree,Abebe2021:MIMO}.
Uncoordinated protocols based on the \ac{CRA} paradigm \cite{casini2007:contention,liva2011:irsa,paolini2015:csa,paolini2015:magazine,clazzer2018:combining,Berioli2016:Modern,Munari2021:age,Valentini2021:coded}, a particular class of grant-free access schemes, ensure high reliability and are currently regarded as candidates for 6G \cite{Mahmood2020:Six} due to their capability of bridging random access with iterative decoding of codes on sparse graphs.

The performance  of \ac{CRA} schemes does not depend only on the \ac{MAC} protocol; it also heavily relies on the effectiveness of the \ac{PHY} layer processing algorithm.
Although part of the literature on \ac{CRA} tends to model the \ac{PHY} layer signal processing (including packet detection, channel estimation, and interference cancellation) as ideal, signal processing in a realistic setting may introduce considerable losses with respect to the performance under idealized conditions. 
Moreover, especially in terrestrial scenarios, the often employed collision or \ac{MPR} channel models tend to be inaccurate, jeopardizing effective system design and optimization \cite{liva2011:irsa,Gha2013:irregular,paolini2015:csa,stefanovic2018:multipacket}. 

In this paper, we investigate \ac{SIC} algorithms for \ac{CRA} in \ac{MMA} applications. In particular, we review and in-depth analyze a low-complexity \ac{SIC} processing proposed in \cite{Sor2018:coded} discussing its vulnerabilities. 
Motivated by this analysis, we propose an innovative massive \ac{MIMO} \ac{SIC} technique able to improve scalability performance. 
The algorithm relies on the fact that, in \ac{CRA}, it is possible to retrieve all packets sent by a user when a subset of them has been decoded. 
Exploiting this knowledge, it is possible to accurately estimate the channel coefficients, which are needed to subtract the interference. 
Due to imperfections of the \ac{SIC} procedure in real scenarios, from now on we adopt the nomenclature \ac{SIS}, to emphasize the non-ideality of this step.

This paper is organized as follows. Section~\ref{sec:preliminary} introduces preliminary concepts, the system model, and some background material. Section~\ref{sec:ImprovingSIC} describes the proposed \ac{SIS} technique along with an analysis which show the improvement of the proposed protocol. Numerical results are shown in Section~\ref{sec:NumericalResults}. Finally, conclusions are drawn in Section~\ref{sec:conclusions}.


\section{Preliminaries and Background}
\label{sec:preliminary}


In this section we define the reference scenario, including the \ac{MAC} protocols and the channel model, also reviewing some physical layer signal processing techniques useful in the next sections.
Throughout the paper, capital and lowercase bold letters denote matrices and vectors, respectively, $(\cdot)^H$ stands for conjugate transposition, $\|\cdot\|$ indicates the Euclidean norm, $\E{\cdot}$ denotes expectation, and $\Var{\cdot}$ is used for variance. 


\subsection{Scenario Definition}\label{subsec:Scenario}

We consider an \ac{MMA} scenario with a very large number $K$ of single-antenna transmitters (also referred to as users or devices), and one receiving \ac{BS} equipped with multiple antennas.
The $K$ users are not all simultaneously active, since they are assumed to wake up unpredictably to transmit one data packet. 
We assume $K_{\mathrm{a}}$ out of the $K$ users are active and the receiver has no prior knowledge of $K_{\mathrm{a}}$. 

We focus on grant-free \ac{MAC} protocols to send uplink data from the active users to the \ac{BS}. 
The schemes of interest belong to the class of \ac{CSA} \cite{paolini2015:csa}, which includes schemes using repetition codes such as \ac{CRDSA} \cite{casini2007:contention} and \acl{IRSA} \cite{liva2011:irsa}. 
Variations on the packets transmission schedule have been proposed in \cite{Valentini2021:coded}.
In this paper we consider \ac{CSA} with repetition codes of a given rate $1/r$ for all users. 
The time is organized in frames, each frame is divided into $N$ slots, and users are frame- and slot-synchronous. Hence, each active user generates $r$ replicas of its packet and transmits them in $r$ slots of the frame. 

The availability of a \ac{BS} with a massive number of antenna elements is a key feature to enable \ac{MPR} at the receiver. 
In this respect, the use of orthogonal pilot sequences (or preambles) represents a simple approach to obtain \ac{MPR} capabilities.
Since in \ac{MMA} $K$ is typically much larger than the number of available pilots $N_\mathrm{P}$, each active user picks one pilot randomly from the set of $N_\mathrm{P}$ preambles, without any coordination with the other active users.
The use of \ac{CSA}-based access and random pilot selection was proposed in \cite{Sor2018:coded}.
Synchronization is achieved exploiting, for example, a beacon transmitted by the \ac{BS} at the beginning of each frame.

Regarding the channel model, we consider a block Rayleigh fading channel with \ac{AWGN}. 
The channel coherence time is assumed equal to the slot duration $T_\mathrm{s}$, which implies statistical independence of the channel coefficients of the same user across different slots. 
We do not consider shadowing effects owing to the assumption of perfect power control.
Coherently with the above-mentioned access protocol and use of orthogonal pilots, each user active in a slot transmits a packet replica composed of one of the $N_\mathrm{P}$ orthogonal pilot sequences, of length  $N_\mathrm{P}$ symbols, concatenated with a payload of length $N_\mathrm{D}$ symbols. Denoting the number of \ac{BS} antennas by $M$, the signal received in a slot may be expressed as $[\M{P}, \M{Y}] \in \mathbb{C}^{M \times (N_\mathrm{P}+N_\mathrm{D})}$ where
\begin{align}\label{eq:P}
    \M{P} = \sum_{k \in \mathcal{A}} \V{h}_k \V{s}(k) + \M{Z}_p \, , \quad \M{Y} = \sum_{k \in \mathcal{A}} \V{h}_k \V{x}(k) + \M{Z} .
\end{align}
In \eqref{eq:P}, $\mathcal{A}$ is the set of users  transmitting a replica in the considered slot, while $\V{h}_k = (h_{k,1}, \dots, h_{k,M})^T \in \mathbb{C}^{M\times 1}$ is the vector of channel coefficients of the $k$-th user. 
The elements of $\V{h}_k$ are modeled as zero-mean, circularly symmetric, complex Gaussian \ac{i.i.d.} random variables, i.e., $h_{k,i} \sim \mathcal{CN}(0, \sigma_\mathrm{h}^2)$ for all $k \in \mathcal{A}$ and $i \in \{1,\dots,M\}$. 
Moreover, $\V{s}(k) \in \mathbb{C}^{1\times N_\mathrm{P}}$ and $\V{x}(k) \in \mathbb{C}^{1\times N_\mathrm{D}}$ are the orthogonal pilot sequence picked by user $k$ in the current slot and the user payload, respectively, both with a unitary average energy per symbol.
Finally, $\M{Z}_p \in \mathbb{C}^{M\times N_\mathrm{P}}$ and $\M{Z} \in \mathbb{C}^{M\times N_\mathrm{D}}$ are matrices whose elements are Gaussian noise samples. The elements of both $\M{Z}_p$ and $\M{Z}$ are \ac{i.i.d.} random variables with distribution $\mathcal{CN}(0, \sigma_\mathrm{n}^2)$.
Due to power control, through the paper we adopt the normalization $\sigma_\mathrm{h}^2 = 1$ for all users' channel coefficients.


\subsection{Channel and Payload Estimation} \label{subsec:MRC}

As mentioned above, the \ac{BS} receives a signal in the form $[\M{P}, \M{Y}]$ in each slot of the frame. The processing can be split into two phases \cite{Sor2018:coded, Valentini2021:coded}. In the first one, the \ac{BS} attempts channel estimation for all possible pilots by computing $\V{\phi}_j\in \mathbb{C}^{M\times 1}$, for all $j \in \{1,\dots,N_{\mathrm{P}}\}$, as
\begingroup
\allowdisplaybreaks
\begin{align}
\label{eq:PhiEstimate}
    \V{\phi}_j &= \frac{\M{P} \,\V{s}_j^{H}}{\| \V{s}_j \|^2} = \sum_{k \in \mathcal{A}^j} \V{h}_k + \V{z}_j
\end{align}
\endgroup
where $\mathcal{A}^j$ is the set of active devices employing pilot $j$ in the current slot, $\V{s}_j \in \mathbb{C}^{1\times N_\mathrm{P}}$ is the $j$-th pilot sequence, and $\V{z}_j \in \mathbb{C}^{M \times 1}$ is a noise vector with \ac{i.i.d.} $\mathcal{CN}(0, \sigma_\mathrm{n}^2 / N_\mathrm{P})$ entries.
Note that in absence of noise, when pilot $j$ is picked by a single user in the current slot, $\V{\phi}_j$ equals the vector of channel coefficients for that user.

In the second phase, the \ac{BS} computes the quantities $\V{f}_j \in \mathbb{C}^{1 \times N_\mathrm{D}}$ and $g_j \in \mathbb{R}$ as
\begingroup
\allowdisplaybreaks
\begin{align}
\label{eq:fComplete}
    \V{f}_j &= \V{\phi}_j^{H} \, \M{Y} \nonumber\\
    &= \sum_{k \in \mathcal{A}^j} \Bigg( \| \V{h}_k \|^2 + \sum_{m \in \mathcal{A}^j \backslash \{k\}} \V{h}_k^H\,\V{h}_m \Bigg) \V{x}(k) \nonumber \\
    &+ \sum_{m \in \mathcal{A} \backslash \mathcal{A}^j} \Bigg( \sum_{k \in \mathcal{A}^j} \V{h}_k^H\,\V{h}_m \Bigg) \V{x}(m) + \V{\tilde{z}}_j
\end{align}
\endgroup
and
\begingroup
\allowdisplaybreaks
\begin{align}
\label{eq:gComplete}
    g_j &= \| \V{\phi}_j \|^2 \nonumber \\
    &= \sum_{k \in \mathcal{A}^j} \Bigg( \| \V{h}_k \|^2 + \sum_{m \in \mathcal{A}^j \backslash \{k\}} \V{h}_k^H\,\V{h}_m \Bigg) + \tilde{n}_j
\end{align}
\endgroup
where $\V{\tilde{z}}_j \in \mathbb{C}^{1 \times N_\mathrm{D}}$ and $\tilde{n}_j$ are noise terms. 
Then, the \ac{BS} attempts estimation of the payload using conventional \ac{MRC} as
\begin{align}
\label{eq:PayloadEst}
    \hat{\V{x}} = \frac{\V{f}_j}{g_j} = \frac{\V{\phi}_j^{H} \, \M{Y}}{\| \V{\phi}_j \|^2}\,.
\end{align}
In the case where a generic user $\ell$ is the only one transmitting with pilot $j$ in a given slot, hereafter referred to as singleton user ($\mathcal{A}^j = \{ \ell \}$), we have $\hat{\V{x}} \approx \V{x}_\ell$. 
Upon successful channel decoding, the packet symbols are stored in a buffer waiting for the \acl{SIS} phase. 
The aim of this iterative processing, that will be explained in detail in the next section, is to subtract the interference of a packet in a slot using the information retrieved in another slot from one of its replicas. 
In fact, whenever a packet is successfully decoded, the \ac{BS} acquires information about the positions of its replicas along with the employed preambles. 
This can be implemented in several ways, e.g., letting this information be a function of the information bits.
This information can be used to subtract interference from a slot and attempt the decoding procedure again.
Here, we separately computed $\V{f}_j$ and $g_j$ for reasons that will be clear in Section~\ref{subsec:NormSquared}.


\section{Analysis of \acl{SIS} Techniques} 
\label{sec:ImprovingSIC}


In this section we present our main contributions. 
We first review in detail a state-of-the-art \ac{SIS} technique for \ac{CSA} with massive \ac{MIMO} \cite{Sor2018:coded}, 
discussing some critical points. Then, we present a theoretical analysis to evaluate the role of interference. 
Motivated by this analysis, we propose a \ac{SIS} algorithm to improve the overall \ac{CSA} scheme performance.


\subsection{Squared-Norm-Based Interference Subtraction}\label{subsec:NormSquared}

Consider the low-complexity \ac{SIS} algorithm, here indicated as \ac{SNB}, proposed in \cite{Sor2018:coded} (also recently exploited in \cite{Valentini2021:coded}). 
It relies on the assumption, whose validity is analyzed and discussed later, that in a massive \ac{MIMO} setting  \eqref{eq:fComplete} and \eqref{eq:gComplete} can be approximated as
\begin{align}
\label{eq:fApprox}
    \V{f}_j &\approx \sum_{k \in \mathcal{A}^j} \| \V{h}_k \|^2 \V{x}(k) + \V{\tilde{z}} \\
\label{eq:gApprox}
    g_j &\approx \sum_{k \in \mathcal{A}^j} \| \V{h}_k \|^2 + \tilde{n}
\end{align}
respectively. 
Assume that we initially compute $\V{f}_{j}$ and $g_j$, $j = 1, \dots, N_\mathrm{P}$, in all slots and that user $\ell$ is successfully decoded in a slot. Then, the above approximations lead naturally to the \ac{SIS} procedure where we update ${\V{f}_j} \leftarrow {\V{f}_j} - \| \V{h}_\ell \|^2\, \V{x}(\ell)$ and ${g_j} \leftarrow {g_j} - \| \V{h}_\ell \|^2$ in all slots 
with replicas of the $\ell$-th user.
In other words, this algorithm subtracts only the main interfering term from \eqref{eq:fComplete} and \eqref{eq:gComplete}.
The update requires to know $\| \V{h}_\ell \|^2$ in the replica slots where, due to the block fading assumption, the channel coefficients are different. 
For this issue, we can use the property that $\| \V{h}_\ell \|^2 /M$ tends to $1$ for large $M$.

Importantly, the approximations \eqref{eq:fApprox} and \eqref{eq:gApprox} are not accurate when the cardinality of $\mathcal{A}$ is large. In fact, since for $m \neq k$
\begin{align}
    \E{\V{h}_k^H\,\V{h}_m} = 0 \, , \quad 
    \Var{\V{h}_k^H\,\V{h}_m} = M
\end{align}
the corresponding interfering terms in \eqref{eq:fComplete} and \eqref{eq:gComplete} may prevent from decoding a user packet even if it is the only one with a specific pilot. 
In the following we analyze this phenomenon by evaluating the probability that a user, being the only one with a specific pilot in a slot, is nevertheless not decoded.

\subsection{Theoretical Analysis of the Interference Effects}
\label{subsec:DecSISNS}

We use the terminology ``logical'' to refer to an idealized setting in which: (i) whenever a user is the only one using a pilot in a given slot it is successfully decoded with probability one;
(ii) channel estimation is perfect so that interference subtraction is ideal.
Hereafter we provide a theoretical analysis of the effects of interference by removing hypotheses (i) and (ii), to understand their impact in a realistic setting. 

Let us consider a situation where $|\mathcal{A}|$ users transmit simultaneously in a slot, $|\mathcal{A}^j|$ of them using pilot $j$.
Assume $|\mathcal{A}^j| - 1$ users from the set $\mathcal{A}^j$ have been successfully decoded in other slots. Then, in the current slot, we can apply \ac{SNB} interference subtraction which, as mentioned above, mitigate but does not eliminate completely the interference. At this point, there is only one undecoded user adopting the $j$-th pilot (singleton). 
To analyze the probability that this user is successfully decoded, we focus on the interfering terms in \eqref{eq:fComplete}. To highlight the effects of the interference we here neglect the noise contribution. Then, from \eqref{eq:fComplete} we can write
\begingroup
\allowdisplaybreaks
\begin{align}
\label{eq:fTerms}
    \V{f}_j &= 
    \sum_{k \in \mathcal{A}^j} \| \V{h}_k \|^2 \, \V{x}(k) + \V{I}_j
\end{align}
\endgroup
with $\V{I}_j = \sum_{i = 1}^{|\mathcal{A}^j| \cdot \left(|\mathcal{A}| - 1 \right)} \V{\xi}_i$.
Each term $\V{\xi}_i$ is expressible as 
$\V{h}_k^H\,\V{h}_m \V{x}$, where $\V{h}_k$ and $\V{h}_m$ are length-$M$ vectors  whose entries are modeled as \ac{i.i.d.} $\mathcal{CN}(0, 1)$ random variables and $\V{x}$ is a length-$N_\mathrm{D}$ payload vector with \ac{i.i.d.} entries. 
It follows that each $\V{\xi}_i$ is a vector whose generic entry fulfills
\begin{align}
\label{eq:XiMeanVar}
    \E{\xi} = 0 \, , \quad \Var{\xi} = M\,.
\end{align}
We can therefore make the approximation 
\begin{align}
    \V{I}_j \approx  \sum_{i = 1}^{|\mathcal{A}^j| \cdot \left(|\mathcal{A}| - 1 \right)} \V{\psi}_i
\end{align}
where $\V{\psi}_i$ are independent random vectors with \ac{i.i.d.} $\mathcal{CN}(0, M)$ entries. 
Due to subtraction of the interference generated by the $|\mathcal{A}^j| - 1$ users decoded in other slots, only one user remains using pilot $j$. Residues of imperfect interference cancellation are incorporated in $\V{I}_j$, yielding a resulting interference term in the form 
\begin{align}
    \tilde{\V{I}}_j \approx \sum_{i = 1}^{N_\mathrm{it}} \V{\psi}_i
\end{align}
where $N_\mathrm{it} = |\mathcal{A}^j| \cdot |\mathcal{A}| - 1$ is the total number of interfering terms.
Performing the estimation as in \eqref{eq:PayloadEst}, we can write
\begin{align}
    \hat{\V{x}}(\ell) = \V{x}(\ell) + \frac{ 1 }{M} \tilde{\V{I}}_j
\end{align}
where the subscript $\ell$ denotes the only remaining user employing pilot $j$.

For a realistic analysis we also consider modulation and channel coding. For example, with a \ac{QPSK} constellation and hard-decision decoding, the symbol error probability is given by 
\begin{align}\label{eq:pe}
    P_\mathrm{e} = \text{erfc}\left(\sqrt{\frac{M}{2 N_\mathrm{it}}}\right) - \frac{1}{4} \text{erfc}^2\left(\sqrt{\frac{M}{2 N_\mathrm{it}}}\right)\,.
\end{align}
Finally, assume an error correcting code with bounded-distance decoding, able to correct up to $t$ errors, and Gray \ac{QPSK} constellation mapping. We can express the probability that decoding of a user packet is unsuccessful in a slot where its $|\mathcal{A}^j| - 1$ pilot-interferers are subtracted and a total of $|\mathcal{A}|$ users were initially allocated in the slot as
\begin{align}
\label{eq:PfailNS}
    P_\mathrm{fail} \approx 1 - \sum_{d = 0}^{t} \binom{N_\mathrm{D}}{d}\, P_\mathrm{e}^d \left( 1 - P_\mathrm{e} \right)^{N_\mathrm{D}-d}
\end{align}
where $N_D$ is the number of payload symbols.

\begin{figure}[t]
    \centering
    \includegraphics[width=0.973\columnwidth]{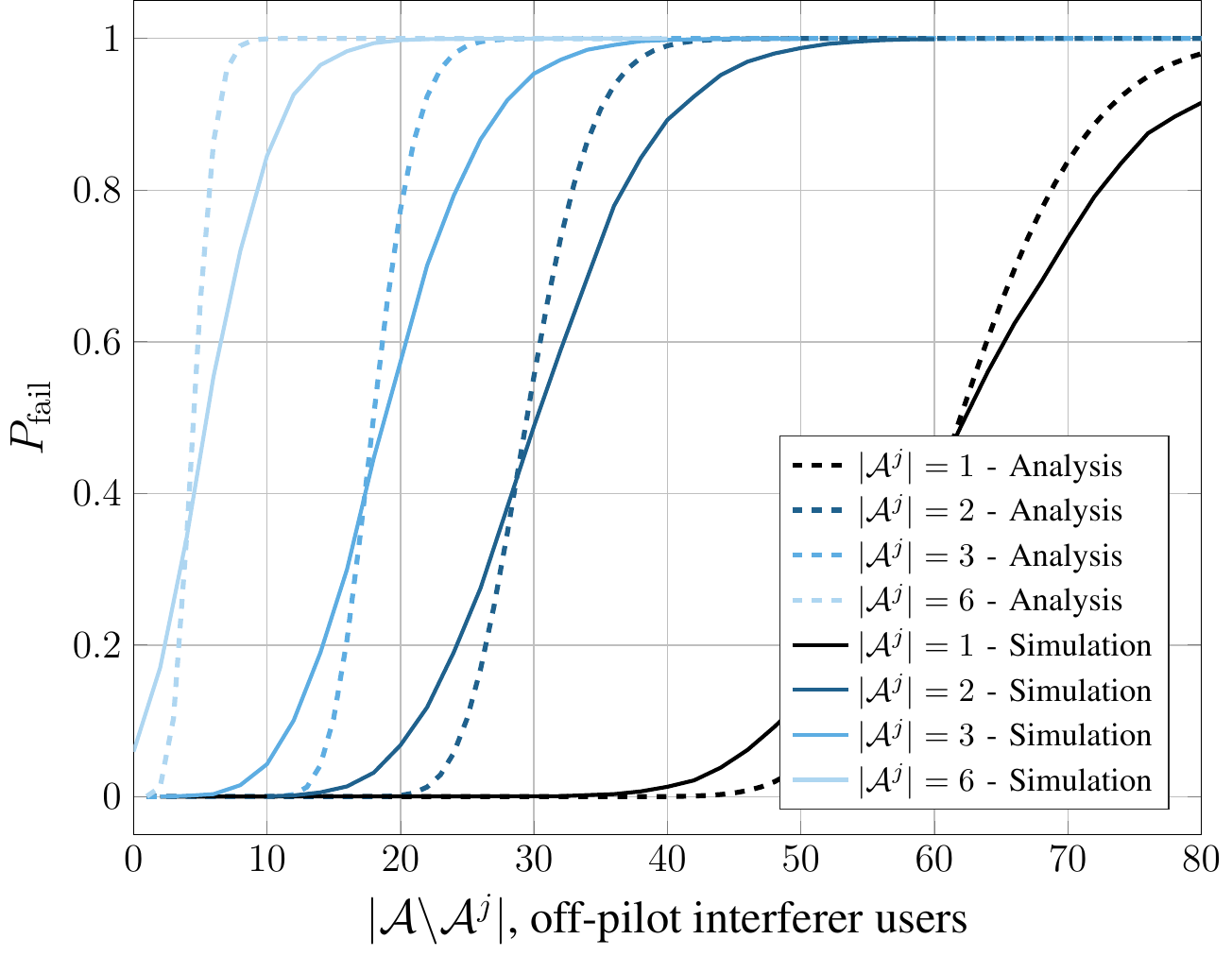}
    \caption{Probability to unsuccessfully decode a singleton user after $|\mathcal{A}^j| - 1$ \ac{SNB} iterations. Comparison between the analytical approximation and the simulation for $N_\mathrm{D} = 256$, $t = 10$, and $M = 256$.}
    \label{fig:InterfNS}
\end{figure}

We report in Fig.~\ref{fig:InterfNS} the analytical approximation \eqref{eq:PfailNS} with $P_\mathrm{e}$ given by \eqref{eq:pe} when $N_\mathrm{D} = 256$, $t = 10$, and $M = 256$. Moreover, we plot the corresponding curves obtained by numerical simulation to validate the derived result. 
Despite the approximations, the analytical results provide a good estimate, in terms of location along the horizontal axis, of the simulated curves.
To improve the system performance in terms of average number of supported users for a given packet error probability of a singleton user, we can increase either the number of \ac{BS} antennas $M$ or the code error correction capability $t$ for fixed $N_\mathrm{D}$ (which decreases the error correcting code rate). %

In the particular case $|\mathcal{A}^j| = 1$, no interference subtraction is necessary and the user experiences the most favorable interference conditions. 
The $|\mathcal{A}^j| = 1$ curve in Fig.~\ref{fig:InterfNS} reveals the actual performance of \ac{MRC} payload estimation in \eqref{eq:PayloadEst} when interferers, using different orthogonal preambles, are captured in the model.
Indeed, this is a major non-ideality, degrading the general performance of \ac{MAC} protocols when a realistic channel model is accounted.
On the other hand, when $|\mathcal{A}^j| > 1$, the estimation deteriorates even more, revealing the non-ideality of the \ac{SIS} procedure. 
Moreover, we point out that, whenever a device using pilot $j$ in the current slot is successfully decoded and \ac{SNB} is performed, the interference on pilots different from $j$ is not mitigated.
This is the critical point of this \ac{SIS} procedure and in Section~\ref{subsec:PayloadAided} we will propose a technique able to overcome this problem.


\subsection{Payload Aided Subtractions}\label{subsec:PayloadAided}

Motivated by the analysis carried out in the previous subsection, we aim at changing the \ac{SIS} algorithm to improve the overall performance.
In repetition-based \ac{CSA}, users send multiple copies of the same payload over the frame.
Hereafter, we refer to the slots used to successfully decode a packet as ``generator'' slots. 

\begin{figure}[t]
    \centering
    \includegraphics[width=1\columnwidth]{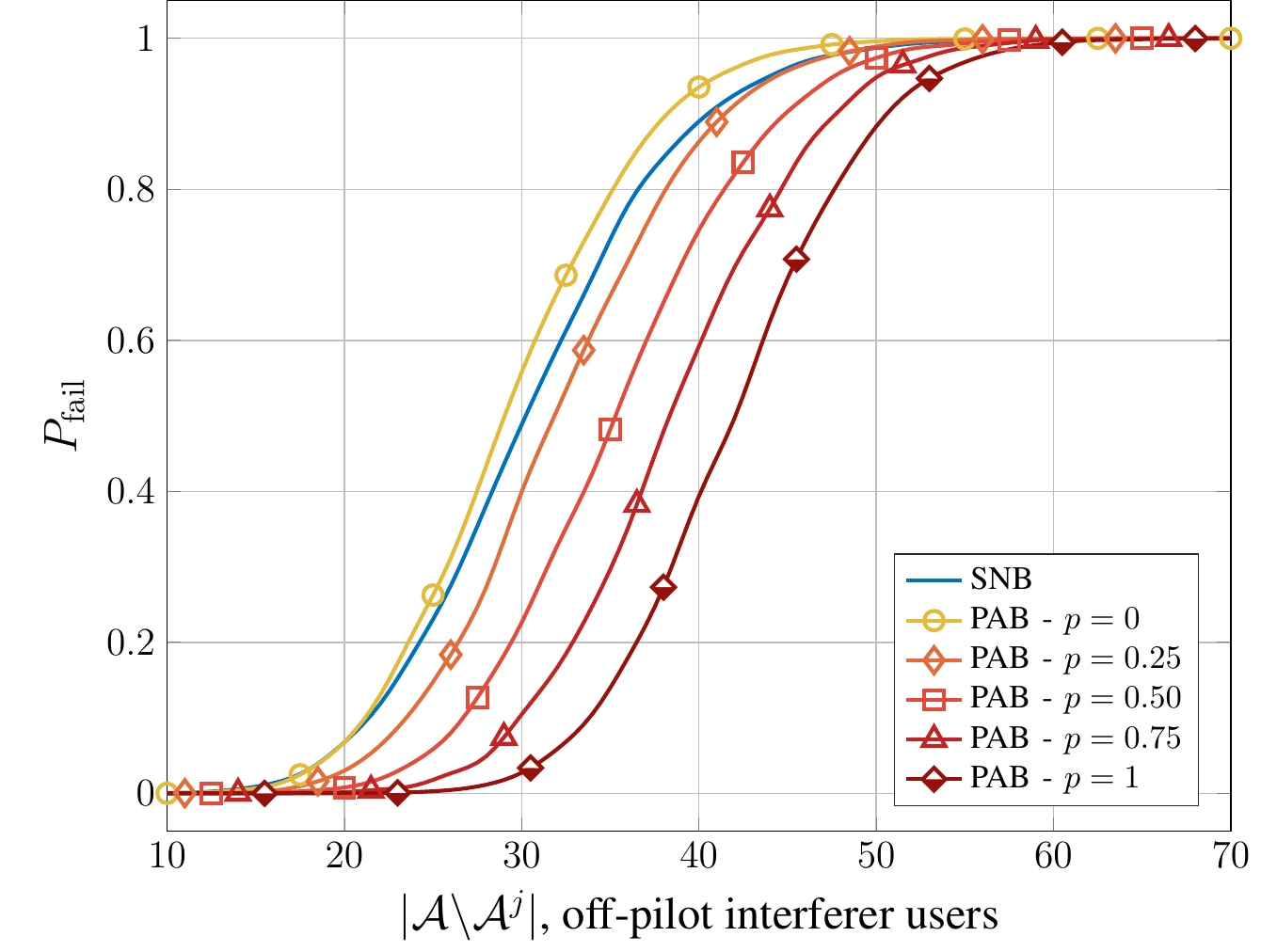}
    \caption{Probability to unsuccessfully decode a singleton user after one \ac{SIS} ($|\mathcal{A}^j| = 2$). Comparison between \ac{SNB} and PAB for $N_\mathrm{D} = 256$, $t = 10$, and $M = 256$.}
    \label{fig:InterfhEst}
\end{figure}

Whenever a user $\ell$ is successfully decoded, the \ac{BS} available information consists of the user's payload of all packets, its channel coefficients in the generator slots (with an accuracy depending on the \ac{AWGN}), the transmission slots, and the chosen pilots in each slot. 
Owing to this information, we can perform the update
\begin{align}\label{eq:PYmod}
    \M{P}^{(i+1)} = \M{P}^{(i)} - \V{h}_\ell \V{s}(\ell) \, , \quad  \M{Y}^{(i+1)} = \M{Y}^{(i)} - \V{h}_\ell \V{x}(\ell)
\end{align}
in the generator slot, where we let $\M{P}^{(0)} = \M{P}$ and $\M{Y}^{(0)} = \M{Y}$. 
Regarding the replica slots, we exploit knowledge of the payload to estimate the channel coefficients as
\begingroup
\allowdisplaybreaks
\begin{align}
\label{eq:hTilde}
    \hat{\V{h}}_\ell &= \frac{\M{Y} \,\V{x}(\ell)^{H}}{\| \V{x}(\ell) \|^2} = \V{h}_\ell + \tilde{\V{h}}_\ell \nonumber\\
    &= \V{h}_\ell + \sum_{k \in \mathcal{A} \backslash \{\ell\}} \V{h}_k \frac{\V{x}(k) \,\V{x}(\ell)^{H}}{\| \V{x}(\ell) \|^2} + \V{z}_\mathrm{h}
\end{align}
\endgroup
where $\V{z}_\mathrm{h}$ is the residual noise and $\M{Y}$ has not been modified yet by other interference subtractions. 
We can derive the statistical properties of the estimation error $\tilde{\V{h}}_\ell$ given that the payload symbols are independent among users, as
\begin{align}
    \E{\tilde{{h}}_{\ell,n}} = 0 \, , \quad  \Var{\tilde{{h}}_{\ell,n}} = \frac{|\mathcal{A}| - 1}{N_\mathrm{D}}
    \label{eq:VarHTilde}
\end{align}
where $n = 1, \dots, M$.
As expected, increasing the number of payload symbols the accuracy of the channel coefficients estimation improves.
We remark that using also knowledge of the preamble to perform channel estimation in slots where we wish to subtract interference may heavily deteriorate the estimation quality due to preamble collisions.
Similarly to \eqref{eq:PYmod}, we can now perform
\begin{align}\label{eq:PYmodTilde}
    \M{P}^{(i+1)} = \M{P}^{(i)} - \hat{\V{h}}_\ell \V{s}(\ell)  \, , \quad 
    \M{Y}^{(i+1)} = \M{Y}^{(i)} - \hat{\V{h}}_\ell \V{x}(\ell)
\end{align}
in the replica slots.
In this \ac{SIS} algorithm, hereafter referred to as \ac{PAB}, each time an update of the matrices $\M{P}$ and $\M{Y}$ has been carried out we re-compute \eqref{eq:PhiEstimate} and \eqref{eq:PayloadEst} for each pilot in the current slot, to check if any other user can be successfully decoded after interference subtraction.
In general, at the step $i = n_\mathrm{up} + n_\mathrm{pa}$ of the \ac{SIS} algorithm, $n_\mathrm{up}$ subtractions are based on uncollided pilots as from \eqref{eq:PYmod}, and $n_\mathrm{pa}$ subtractions are based on payload aided channel coefficients estimation as from \eqref{eq:PYmodTilde}.

Fig.~\ref{fig:InterfhEst} illustrates the results of a variation of the experiment described in Section~\ref{subsec:DecSISNS} for the two \ac{SIS} techniques discussed in this paper and for $|\mathcal{A}^j| = 2$. 
More specifically, we assume that a fraction $0 \le p \le 1$ of users in the set $\mathcal{A} \backslash \mathcal{A}^j$ have been successfully decoded and subtracted. For them, we consider the worst case scenario ($n_\mathrm{up} = 0$) where the \ac{SIS} is performed using \eqref{eq:PYmodTilde}.
As expected, the \ac{PAB} performance improves as $p$ increases. On the other hand, \ac{SNB} is not influenced by $p$. As in a real scenario we have $n_\mathrm{up} > 0$, the \ac{PAB} technique is expected to outperform the \ac{SNB} one; this is confirmed by the numerical results presented in the next section.


\section{Performance Evaluation}
\label{sec:NumericalResults}


\subsection{Simulation Setup}
\label{subsec:SimSetUp}

We present numerical results about the \ac{SIS} techniques discussed in the previous sections using different \ac{MAC} protocols.
We consider a system where users transmit payloads encoded with an $(n, k, t)$ narrow-sense binary \ac{BCH} code. A \ac{CRC} code is also used to validate decoded packets and avoid that the \ac{SIS} procedure adds interference instead of subtracting it. 
Zero padding the \ac{BCH} codeword with a final bit, we can map encoded bits onto a \ac{QPSK} constellation with Gray mapping, obtaining $N_\mathrm{D}$ symbols per codeword. The \ac{QPSK} symbol energy is normalized to $1$.
Simulations have been carried out with symbol rate $B_\mathrm{s} = 1$~Msps, $M = 256$ \ac{BS} antennas, and $\sigma_\mathrm{n}^2 = 0.1$.
We also impose a maximum latency constraint $\Omega = 50$~ms, leading to a number of slots per frame $N$ equal to \cite{Valentini2021:coded}
\begin{align}
\label{eq:NSlot}
    N = 
    \left\lfloor \frac{\Omega \, B_\mathrm{s}}{2 \, (N_\mathrm{P} + N_\mathrm{D}) } \right\rfloor
\end{align}
where the number of orthogonal pilot symbols, $N_\mathrm{P}$, equals the total number of available pilot sequences.
These sequences are constructed using Hadamard matrices.


\subsection{Numerical Results}
\label{subsec:NumRes}

\begin{figure}[t]
    \centering
    \includegraphics[width=0.99\columnwidth]{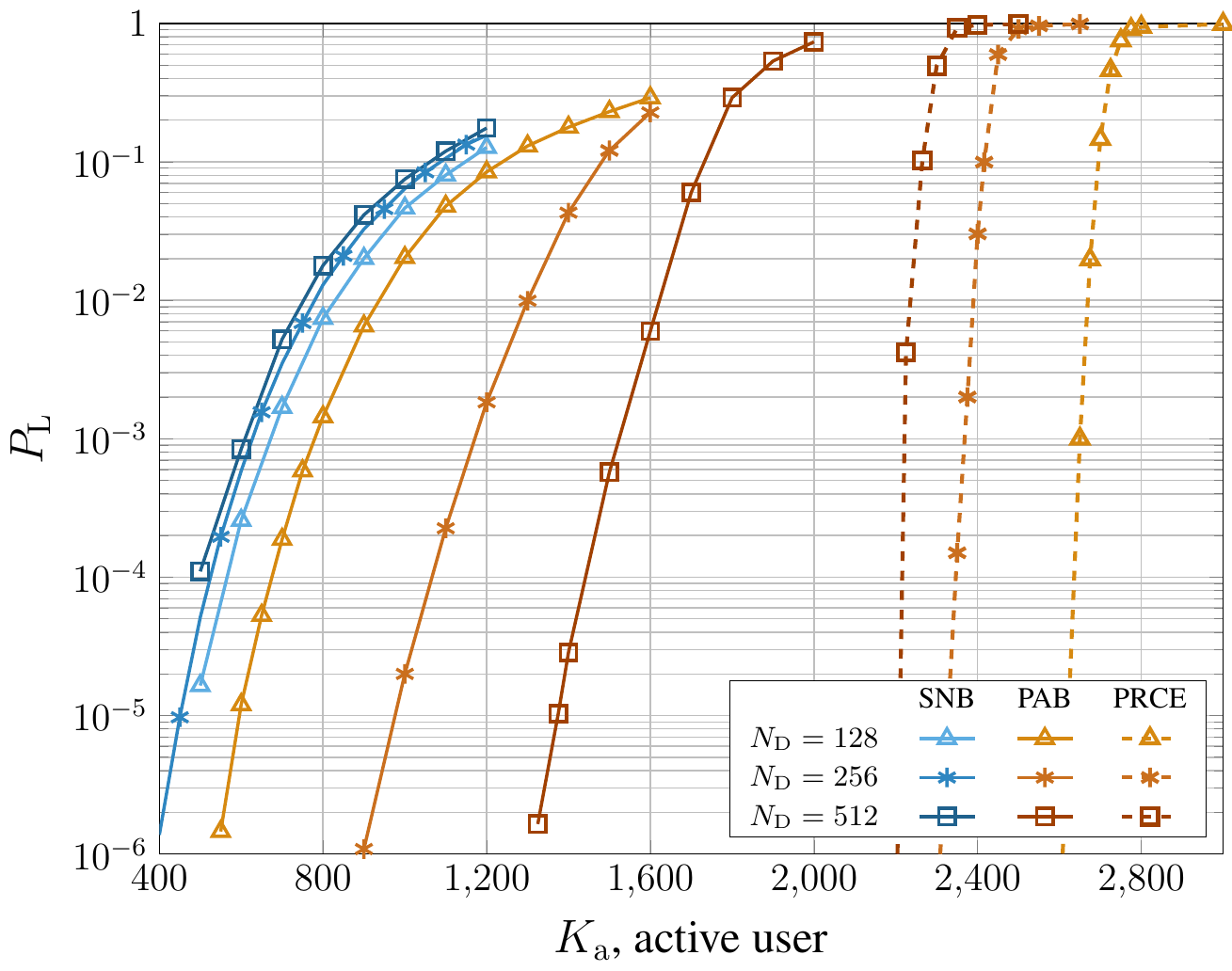}
    \caption{Packet loss rate values of schemes characterized by different \ac{SIS} techniques and payload sizes $N_\mathrm{D}$. Baseline \ac{MAC} with $N_\mathrm{P} = 64$, $N = 78$. Comparison between the \acs{SNB}, the proposed \acs{PAB} and the ideal \acs{SIC} case denoted as \acs{PRCE}.}
    \label{fig:diffSymb}
\end{figure}

We compare \ac{SIS} techniques in terms of \acl{PLR} $P_\mathrm{L}$ for a given number $K_\mathrm{a}$ of active users in the frame. 
Regarding the \ac{MAC} protocol, we adopt a standard repetition-based \ac{CSA} protocol with a constant number $r$ of replicas per packet \cite{casini2007:contention}, referred to in the following as the ``baseline \ac{MAC}''. 
As a variation of the baseline protocol, we also show results for a second \ac{MAC} protocol, namely, the recently proposed repetition-based \ac{CSA} with intra-frame \ac{SC} scheduling and \ac{ACK} messages \cite{Valentini2021:coded}.
As a reference upper bound, we report the performance of a logical simulation. 
In this idealized setting a user transmitting alone in a ``resource'' (slot-pilot pair) uncollided is successfully decoded with probability one. 
Finally, as a second upper bound for the proposed scheme, we consider also a simulation which performs the \ac{PAB} processing under the assumption that the subtractions are perfect (ideal \ac{SIC}). 
In this scheme, denoted as \acf{PRCE}, the performance is limited by the payload estimation \eqref{eq:PayloadEst}. 

In Fig.~\ref{fig:diffSymb} we report the \ac{PLR} varying the symbol payload size $N_\mathrm{D}$ while keeping the rate of the \ac{BCH} code constant, for the \ac{SNB}, \ac{PAB}, and \ac{PRCE} techniques. To be precise, for $N_\mathrm{D} \in \{ 128, 256, 512 \}$ the corresponding \ac{BCH} codes are $(255, 207, 6)$, $(511, 421, 10)$, and $(1023, 843, 18)$. In this particular example, we adopt the baseline \ac{MAC} fixing $N = 78$ and $N_\mathrm{P} = 64$ in order to show only the influence of $N_\mathrm{D}$ in the \ac{SIS} processing.
Looking at the curves of the \ac{SNB} processing, we observe that the performance slightly degrades when $N_\mathrm{D}$ increases. The same behavior can be observed for \ac{PRCE}. On the other hand, the \ac{PAB} improves when $N_\mathrm{D}$ increases, as expected from \eqref{eq:VarHTilde}. In addition, from Fig.~\ref{fig:diffSymb} we can see that the gap between the \ac{PRCE} and the \ac{PAB} reduces, highlighting the effectiveness of the proposed technique in a complete scenario which accounts for both the \ac{PHY} and \ac{MAC} layer.

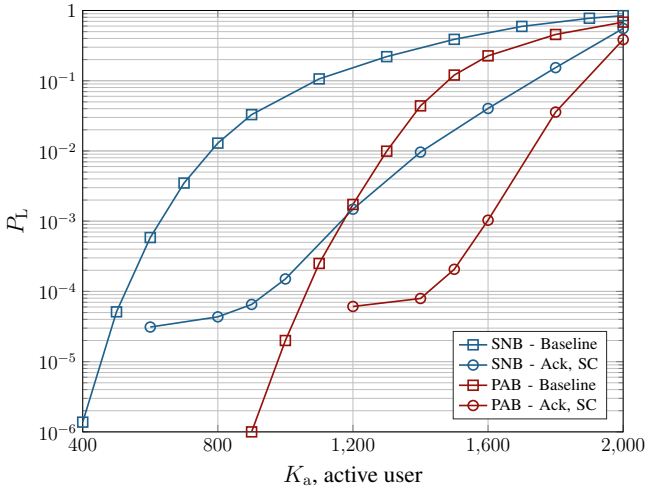
\begin{figure}[t]
    \centering
    \resizebox{0.99\columnwidth}{!}{
%
%
\definecolor{mycolorGreen}{rgb}{0.28627,0.68627,0.30588}%
\definecolor{mycolorBlue}{rgb}{0.29020,0.31373,0.69020}%
\definecolor{graphOrange}{rgb}{0.8902  ,  0.5804  ,  0.2275}%
\definecolor{graphPurple}{rgb}{0.49412,0.18431,0.55686}%
\definecolor{mycolorRed}{rgb}{0.7608, 0.2588, 0.1255}
\definecolor{newDeepBlue}{rgb}{0.1216, 0.3804, 0.5529}%
\definecolor{graphRed}{rgb}{0.76078,0.13725,0.13725}%
\definecolor{graphDarkRed}{rgb}{0.5882, 0.0667, 0.0353}%
\begin{tikzpicture}

\begin{axis}[%
width=4.521in,
height=3.552in,
at={(0.758in,0.495in)},
scale only axis,
xmin=400,
xmax=2000,
xlabel style={font=\color{white!15!black}, font=\Large},
xlabel={$K_\mathrm{a}$, active user},
xtick distance = 400,
ymode=log,
ymin=1e-06,
ymax=1,
yminorticks=true,
yticklabels = {0, $10^{-6}$, $10^{-5}$, $10^{-4}$, $10^{-3}$, $10^{-2}$, $10^{-1}$, $1$},
ylabel style={font=\color{white!15!black}, font=\Large},
ylabel={$P_\mathrm{L}$},
tick label style={black, semithick, font=\large},
axis background/.style={fill=white},
xmajorgrids,
ymajorgrids,
yminorgrids,
legend style={at={(0.97,0.03)}, anchor=south east, legend cell align=left, align=left, draw=white!15!black}
]


\addplot [color=newDeepBlue, line width=1.0pt, mark size=3.0pt, mark=square, mark options={solid, newDeepBlue}]
  table[row sep=crcr]{%
400	1.37499999997015e-06\\
500	5.1200000000029e-05\\
600	0.000584833333333368\\
700	0.00349271428571429\\
800	0.0129305\\
900	0.0329018888888889\\
1100	0.106169696969697\\
1300	0.220287179487179\\
1500	0.388345185185185\\
1700	0.593207843137255\\
1900	0.775281871345029\\
2000	0.843805555555556\\
};
\addlegendentry{\acs{SNB} - Baseline}

\addplot [color=newDeepBlue, line width=1.0pt, mark size=3.0pt, mark=o, mark options={solid, newDeepBlue}]
  table[row sep=crcr]{%
600	3.10884353741381e-05\\
800	4.34183673468924e-05\\
900	6.51700680271983e-05\\
1000	0.000150408163265281\\
1200	0.00148159863945574\\
1400	0.00962956268221571\\
1600	0.0402749999999999\\
1800	0.153526666666667\\
2000	0.55494\\
2100	0.746815238095238\\
2300	0.906718260869565\\
};
\addlegendentry{\acs{SNB} - Ack, \ac{SC}}




\addplot [color=graphDarkRed, line width=1.0pt, mark size=3.0pt, mark=square, mark options={solid, graphDarkRed}]
  table[row sep=crcr]{%
900     1.0e-06\\
1000	2.000000000002e-05\\
1100	0.000249999999999972\\
1200	0.00173166666666669\\
1300	0.00992692307692311\\
1400	0.0438757142857142\\
1500	0.120416666666667\\
1600	0.22606125\\
1800	0.455513333333333\\
2000	0.682035\\
};
\addlegendentry{\acs{PAB} - Baseline}

\addplot [color=graphDarkRed, line width=1.0pt, mark size=3.0pt, mark=o, mark options={solid, graphDarkRed}]
  table[row sep=crcr]{%
1200 6.1e-5\\
1400	7.92857142857262e-05\\
1500	0.000206400000000051\\
1600	0.00103249999999999\\
1800	0.0357711111111111\\
2000	0.385575555555556\\
2200	0.771812929292929\\
2400	0.90448375\\
2600	0.947571324786325\\
};
\addlegendentry{\acs{PAB} - Ack, \ac{SC}}

\end{axis}
\end{tikzpicture}%
    }
    \caption{Packet loss rate comparison between \ac{SNB} and the proposed \ac{PAB}, maximum latency $\Omega = 50$~ms, $N_\mathrm{P} = 64$, $N = 78$, $N_\mathrm{D} = 256$, for \ac{MAC} protocols with or without \acf{SC}.}
    \label{fig:perfZoom}
\end{figure}

In Fig.~\ref{fig:perfZoom} we plot the comparison between the \ac{SNB} and the \ac{PAB} imposing a maximum latency $\Omega = 50$~ms, $N_\mathrm{P} = 64$ available pilots, repetition rate $r = 3$, a $(511, 421, 10)$ \ac{BCH} code, using both the baseline \ac{MAC} protocol and the \ac{SC} with \acp{ACK} \cite{Valentini2021:coded}.
From these curves we observe that \ac{PAB} remarkably improves the performance in comparison to \ac{SNB}. For example, at $P_\mathrm{L} = 10^{-4}$ \ac{SNB} can support  $K_\mathrm{a} \approx 550$ users, while \ac{PAB} more than twice $K_\mathrm{a} \approx 1100$.
 
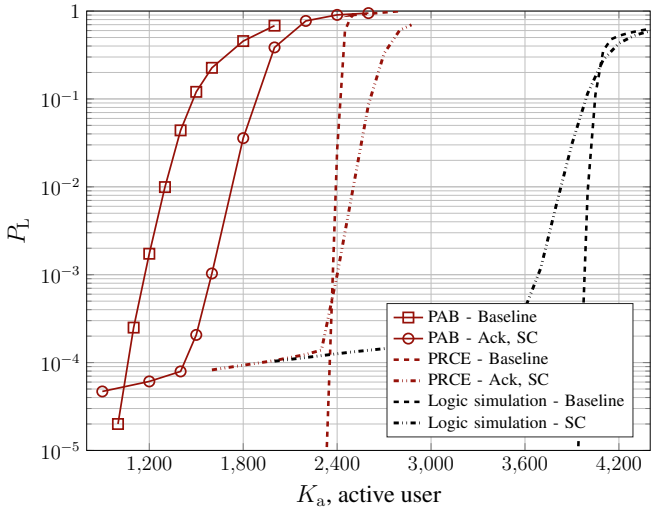
\begin{figure}[t]
    \centering
    \resizebox{0.99\columnwidth}{!}{
%
%
\definecolor{mycolorGreen}{rgb}{0.28627,0.68627,0.30588}%
\definecolor{mycolorBlue}{rgb}{0.29020,0.31373,0.69020}%
\definecolor{graphOrange}{rgb}{0.8902  ,  0.5804  ,  0.2275}%
\definecolor{graphPurple}{rgb}{0.49412,0.18431,0.55686}%
\definecolor{mycolorRed}{rgb}{0.7608, 0.2588, 0.1255}
\definecolor{newDeepBlue}{rgb}{0.1216, 0.3804, 0.5529}%
\definecolor{graphRed}{rgb}{0.76078,0.13725,0.13725}%
\definecolor{graphDarkRed}{rgb}{0.5882, 0.0667, 0.0353}%
\begin{tikzpicture}

\begin{axis}[%
width=4.521in,
height=3.552in,
at={(0.758in,0.495in)},
scale only axis,
xmin=800,
xmax=4400,
xlabel style={font=\color{white!15!black}, font=\Large},
xlabel={$K_\mathrm{a}$, active user},
xtick distance = 600,
ymode=log,
ymin=1e-05,
ymax=1,
yminorticks=true,
yticklabels = {0, $10^{-5}$, $10^{-4}$, $10^{-3}$, $10^{-2}$, $10^{-1}$, $1$}, 
ylabel style={font=\color{white!15!black}, font=\Large},
ylabel={$P_\mathrm{L}$},
tick label style={black, semithick, font=\large},
axis background/.style={fill=white},
xmajorgrids,
ymajorgrids,
yminorgrids,
legend style={at={(0.97,0.03)}, anchor=south east, legend cell align=left, align=left, draw=white!15!black}
]

\addplot [color=graphDarkRed, line width=1.0pt, mark size=3.0pt, mark=square, mark options={solid, graphDarkRed}]
  table[row sep=crcr]{%
1000	2.000000000002e-05\\
1100	0.000249999999999972\\
1200	0.00173166666666669\\
1300	0.00992692307692311\\
1400	0.0438757142857142\\
1500	0.120416666666667\\
1600	0.22606125\\
1800	0.455513333333333\\
2000	0.682035\\
};
\addlegendentry{\acs{PAB} - Baseline}

\addplot [color=graphDarkRed, line width=1.0pt, mark size=3.0pt, mark=o, mark options={solid, graphDarkRed}]
  table[row sep=crcr]{%
900	4.69135802468879e-05\\
1200 6.1e-5\\
1400	7.92857142857262e-05\\
1500	0.000206400000000051\\
1600	0.00103249999999999\\
1800	0.0357711111111111\\
2000	0.385575555555556\\
2200	0.771812929292929\\
2400	0.90448375\\
2600	0.947571324786325\\
};
\addlegendentry{\acs{PAB} - Ack, \ac{SC}}


\addplot [color=graphDarkRed, line width=1.5pt, dashed]
  table[row sep=crcr]{%
2300	1.93423597694142e-07\\
2400	0.026155119047619\\
2450	0.596342857142857\\
2475	0.758988282828283\\
2500	0.9124992\\
2525	0.931348910891089\\
2500	0.8995544\\
2800	0.99671\\
};
\addlegendentry{PRCE - Baseline}

\addplot [color=graphDarkRed, dashdotdotted, line width=1.7pt]
  table[row sep=crcr]{%
1600	8.24999999999854e-05\\
2000	0.00010542857142859\\
2200	0.000123636363636417\\
2300	1.38888888889161e-04\\
2400	0.000994666666666699\\
2600	0.0860232692307692\\
2700	0.329755555555556\\
2800	0.606792857142857\\
2875	0.697349565217391\\
};
\addlegendentry{PRCE - Ack, SC}


\addplot [color=black, line width=1.5pt, dashed]
  table[row sep=crcr]{%
2000	5.71428571127797e-08\\
2800	6.12244898112735e-08\\
3400	8.0580177308498e-08\\
3900	1.12399016538589e-07\\
4000	0.00867966374269002\\
4050	0.112083245149912\\
4100	0.355170975609756\\
4150	0.477035180722892\\
4200	0.523117589285714\\
4250	0.5545505\\
4300	0.582280726744186\\
4350	0.606009224137931\\
4400	0.626930795454545\\
};
\addlegendentry{Logic simulation - Baseline}

\addplot [color=black, dashdotdotted, line width=1.7pt]
  table[row sep=crcr]{%
2000	0.000103569500000011\\
2500	0.000132292400000011\\
3000	0.000163434666666684\\
3500	0.000209000000000015\\
3600	0.00041295138888886\\
3700	0.0011769594594595\\
3800	0.0062210526315789\\
3900	0.0307325236167342\\
4000	0.1153675\\
4100	0.279406129653402\\
4200	0.42978242481203\\
4300	0.52929541003672\\
4400	0.594112290669856\\
4500	0.639920116959064\\
};
\addlegendentry{Logic simulation - \acs{SC}}

\end{axis}
\end{tikzpicture}%
    }
    \caption{
    Packet loss rate comparison between the proposed \ac{PAB} and its bounds given by \ac{PRCE} and logic simulation.
    Maximum latency $\Omega = 50$~ms, $N_\mathrm{P} = 64$, $N = 78$, $N_\mathrm{D} = 256$, for \ac{MAC} protocols with or without \acf{SC}.}
    \label{fig:perfAll}
\end{figure}

In Fig.~\ref{fig:perfAll} we extend Fig.~\ref{fig:perfZoom} aiming to point out the gap between a real system represented by \ac{PAB} and two idealized schemes, the \ac{PRCE} and the logic one. As anticipated in Fig.~\ref{fig:diffSymb}, the distance between the \ac{PAB} curve and the \ac{PRCE} is mainly due to the channel estimation imperfections. On the other hand, the gap between the \ac{PRCE} and the logical simulation is a consequence of the payload estimation non-ideality addressed in Section~\ref{subsec:DecSISNS}. This plot reveals how neglecting the \ac{PHY} layer processing in real scenario may lead to wrong conclusions and optimizations.

\section{Conclusions}\label{sec:conclusions}


Interference poses a serious challenge for next generation grant-free \acl{MMA}. In this paper we provided an in-depth analysis of this problem for \ac{CRA} schemes and proposed a massive \ac{MIMO} interference subtraction processing able to ameliorate the scalability of state-of-the-art schemes, in the presence of reliability and latency constraints.
For example, with a maximum latency of $50$~ms and a target packet loss rate $P_\mathrm{L} = 10^{-4}$, the proposed scheme is able to double the number of served users compared to the baseline.
We also emphasized a large gap between results obtained under idealized and realistic condition, revealing how system design and analysis relying on collision-like channels may turn inaccurate.


\section*{Acknowledgements}

The Authors would like to thank the anonymous Reviewers for comments and suggestions. 
This work has been carried out in the framework of the CNIT National Laboratory WiLab and the WiLab-Huawei Joint Innovation Center.





\end{document}